# FastHDRNet: A new efficient method for SDR-to-HDR Translation

Tian Siyuan[1], Wang Hao[1], Rong Yiren[1], Wang Junhao[1], Dai Renjie[1]and He Zhengxiao[1(✉)]

[1] Tongji University, Caoan Road 4800, Shanghai 201804, China
1950095@tongji.edu.cn

**Abstract.** Modern displays nowadays possess the capability to render video content with a high dynamic range (HDR) and an extensive color gamut .However, the majority of available resources are still in standard dynamic range (SDR). Therefore, we need to identify an effective methodology for this objective.The existing deep neural networks (DNN) based SDR to HDR conversion methods outperforms conventional methods, but they are either too large to implement or generate some terrible artifacts. We propose a neural network for SDR to HDR conversion, termed "FastHDRNet". This network includes two parts, Adaptive Universal Color Transformation (AUCT) and Local Enhancement (LE). The architecture is designed as a lightweight network that utilizes global statistics and local information with super high efficiency. After the experiment, we find that our proposed method achieves state-of-the-art performance in both quantitative comparisons and visual quality with a lightweight structure and a enhanced infer speed.

## 1    Introduction

In recent years, the transition towards High Dynamic Range (HDR) technology has significantly enhanced the visual quality of television and film content, evolving from Standard Definition through Full High Definition to Ultra-High Definition. With its expansive color gamut and superior dynamic range, HDR technology surpasses Standard Dynamic Range (SDR) capabilities, presenting visuals that mirror real-life more closely. Standards such as DCI-P3 [1] and BT.2020 [2] have been crucial in establishing HDR display qualities, facilitating displays to reach over 2000 nits of peak brightness, in contrast to SDR's limitations.

Despite display technology advancements, HDR content production and accessibility lags behind, creating a need for proficient SDR-to-HDR conversion methods. Earlier conversion techniques, which relied on inverse tone mapping based on image statistics, have been largely replaced by sophisticated deep neural networks

---

(DNNs) strategies [3, 4, 5, 6, 7], including convolutional neural networks (CNNs). These methods excel by learning the correlation between SDR and HDR through paired dataset training, although the large-scale DNN deployment faces integration challenges in consumer devices due to computational complexity.

Newer research has directed efforts towards crafting algorithms that offer SDR to HDR conversion with less computational demand. This includes developing lighter networks, which retains quality conversion while being more suited for integration into daily use devices like TVs and AR/VR equipment. Furthermore, recognizing that SDR and HDR content are processed differently despite originating from identical raw files, conversion tasks now also consider dynamic range, color gamuts, and bit depths. Among these innovations, Efficient-HDRTV [8] promises to deliver quality HDRTV content with significantly less computational overhead.

Additionally, applying transformers in low-level vision tasks suggests a method for enhancing SDR-to-HDR conversion, particularly for translating both low and high-frequency detail information. Despite direct application challenges, adjusting attention mechanisms to selectively enhance specific frequency features can refine the conversion process.

The rapid evolution of display technologies and escalating HDR content demand call for continued research and development in SDR-to-HDR conversion. Progress in DNNs and the introduction of more efficient algorithms signifys notable advances, offering viable solutions to close the gap between HDR display capabilities and content availability. Expectedly, these innovations will not only improve consumer visual experiences but also promote HDR technology's broader adoption in media and entertainment.

In summary, our contributions are three-fold:

- We propose a lightweight and efficient SDR-HDR method that achieves state-of-the-art performance.
- We experiment with new normalization method to make this network more robustly.
- Our method has the fastest reference time in all the algorithms mentioned.

As you can see from **Fig. 1**, our model achieves high quantitative performance with a very low parameter numbers.

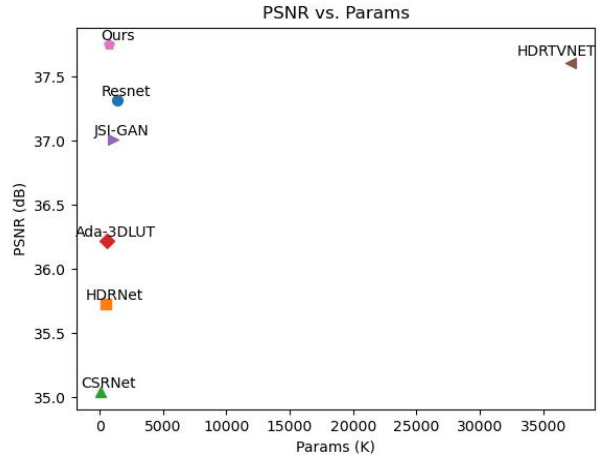

**Fig. 1.** Comparison of number of parameters among models.

## 2    Related Work

The adoption of HDR in television content, as per the specifications of the Rec.2100 standard [9], marks a crucial development beyond the earlier Rec.709 standard [10] that established the framework for SDR displays. Distinguished by an enhanced color depth (10 or 12 bits per pixel compared to the 8 bits per pixel of SDR), an expanded color gamut, and the employment of sophisticated optical-electronic transfer functions (OETF) [9], HDR technology significantly augments the encoding of media content. It provides a richer palette of colors and a more lifelike viewing experience by capturing intricate details in both bright and dark areas. This study introduces a pioneering deep learning model for converting SDR content into HDR, aiming at capitalize on the benefits of HDR technology to improve television production quality.

### 2.1    Background

The Rec.2100 standard [9] defines HDR imagery in the pixel domain, facilitating a more precise representation of physical brightness via a luminance map in the linear domain [5]. This represents a substantial leap from SDR processing, which involves dynamic range clipping, non-linear mapping with a camera response function, and quantization [11, 12]. These processes contribute significantly to creating a more engaging viewing experience.

### 2.2    SDR-to-HDR Translation

The translation from SDR to HDR is necessitated by the inherent loss of information in SDR media when compared to HDR. Initial methodologies focused on multi-purpose CNN [13] for HDR reconstruction. Further developments introduced techniques such as Deep SRITM [14] for decomposing SDR images into detailed components for contrast enhancement, and the JSIGAN [15] for learning pixel-wise filters for local contrast enhancement using an adversarial framework. Chen et al. [5] developed the HDRTVNet, designed to reconstruct HDR through adaptive global color mapping, local enhancement, and highlight generation, further supported by the HDRTV1K benchmark dataset. Additional innovations include the frequency-aware modulation network by Xu et al. [16] and a two-stage scheme by He et al. [17] for improving local region reconstruction quality.

Despite these advances, earlier CNN-based approaches [5, 14, 15] overlooked the significance of non-local dependencies among pixels. This research introduces an advanced attention mechanism to model long-range feature correlations, enhancing the quality of SDR-to-HDR translation by leveraging the capabilities of DNNs [18-22] and the Vision Transformer [23].

## 2.3 Vision Transformer

The advent of attention-based Transformer models [23], initially prominent within the realm of natural language processing (NLP), has progressively infiltrated various domains of computer vision, such as image recognition [24], semantic segmentation [25], and object detection [26]. Recent research has ventured into the application of Transformers for low-level vision tasks [5-7], heralding models like SwinIR [26] that leverages local attention mechanisms, and Restormer [22], which employs attention across channel dimensions to facilitate efficient image restoration.

In light of the shortcomings associated with traditional SDR-to-HDR conversion methodologies [5, 8], this study introduces a comprehensive strategy that synergizes the precision of DNNs with the broad analysis capabilities of Vision Transformers. This strategy encompasses mastering complex non-linear mappings, utilizing U-Net [41] architectures for meticulous detail enhancement [18], and adopting sophisticated attention mechanisms to grasp long-range dependencies, thereby crafting an encompassing SDR-to-HDR translation solution.

However, it's noteworthy that these Transformer-based approaches may exhibit slower inference speeds compared to CNN-based alternatives, incurring higher computational demands. Such attributes potentially limit their practical utility in real-world applications.

## 3 Method

The transition of television content from SDR to HDR marks a pivotal enhancement in visual quality. In response to the complexities of this transition, we present FastHDRNet, a novel framework consisting of two deep neural networks designed for specific stages of the SDR-to-HDR conversion. Initially, a pixel-independent operation addresses both average and peak brightness, followed by a region-independent refinement crucial for accurate HDR transformation. This strategy notably reduces artifact occurrences, offering a more seamless conversion. An indepth overview of FastHDRNet's core components, as illustrated in **Fig. 2**, will elucidate the advanced mechanisms underpinning this conversion method.

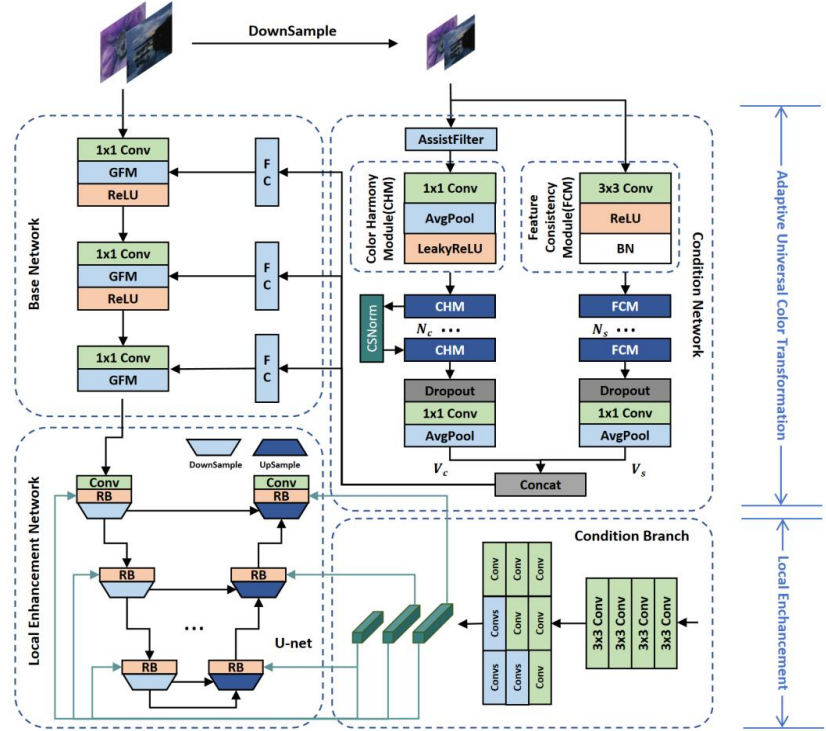

**Fig. 2.** The architecture of the proposed FastHDRNet method.

### 3.1 Adaptive Universal Color Transformation (AUCT)

The AUCT framework is engineered to adeptly execute image-specific color mapping, seamlessly converting the color profile of an SDR image into its HDR equivalent. Structurally, AUCT is composed of two pivotal components: a foundational network and a conditioning network. These components collaboratively underpin the process of global color adjustment, ensuring a nuanced translation of color spaces from SDR to HDR contexts.

**Base Network**

At the heart of the AUCT framework lies the base network, tasked with implementing global operations at the pixel level throughout the image. This operation is formally described as:

$$M_B(x, y) = f(M_S(x, y)), \quad \forall (x, y) \in M_S \tag{1}$$

where $M_S$ denotes the input SDR image, $(x, y)$ specifies the pixel coordinates, and $M_B$ symbolizes the output generated by the base network. Inspired by the CSRNet [19], our base network is structured as a fully convolutional network, leveraging $1 \times 1$ convolutions paired with Rectified Linear Unit (ReLU) activation functions to accomplish global mapping. This setup is succinctly represented as:

$$M_B = Conv_{1\times1} \circ (ReLU \circ Conv_{1\times1})^{N_l-1}(M_S) \tag{2}$$

where $N_l$ denotes the number of convolutional layers, each equipped with $1 \times 1$ filters, followed by $N_l - 1$ ReLU activations. This configuration enables the base network to process an 8-bit SDR image and output an HDR representation with a depth ranging from 10 to 16 bits. Remarkably, the base network's ability to emulate a 3D lookup table (3D LUT) functionality with a reduced parameter set, as opposed to direct 3D LUT learning, showcases its efficacy in color mapping. For additional results and a more detailed exposition of the base network's performance, readers are directed to the supplementary material.

### Condition Network

Global Priors are definitely important in modulation of the base network. In this part, we only care about the global luminance and structure of the whole picture. So we devise two kinds of blocks, Color Harmony Module (CHM) and Feature Consistency Module (FCM). We would like to take color conformity and structural similarity into consideration. Different images have different holistic colors, specific structures, and global image statistics. To fully extract priors, we propose a Condition Network CHM and FCM. CSRNet [12] extracts the global priors using $N_k \times N_k(N_k > 1)$ filters for image retouching. Furthermore, AGCM [5] uses $1 \times 1$ filters to focus on extracting global color priors for the SDR-to-HDR task. Thus, inspired by these methods, we introduce the condition network branch for SDR-to-HDR. Unlike AGCM, we perform a guided filtering [27] of the low-resolution frame before feeding it to the convolution layers. This condition network can simultaneously extract high-frequency and low-frequency information. Finally, we concatenate the output of these two branches and broadcast it to the base network for modulation.

### Color Harmony Module (CHM)

Within the FastHDRNet architecture, a pivotal element named the CHM significantly contributes to the enhancement of the adaptive global color mapping procedure. This module incorporates a meticulously arranged series of layers, each playing a part in a tailored pipeline aimed at refining the color transition from SDR to HDR formats. The functionality of the CHM is articulated through a mathematical framework as follows:

$$CHM(\cdot) = CSNorm \circ LReLU \circ avgpool \circ Conv_{1\times1}(\cdot) \tag{3}$$

CSNorm , delineated in **Fig. 3**, represents a sophisticated approach to lightness adaptation, employing channel-selective normalization [28]. This methodology has demonstrated remarkable efficiency and resilience across various lighting scenarios. It incorporates the Leaky ReLU as its activation function, utilizes an average pooling operation, and employs a convolution layer equipped with 1×1 filters. Through this structured sequence, the CSNorm processes the input via the CHM to finely tune color harmony, ensuring an optimal adjustment tailored to the input's characteristics.

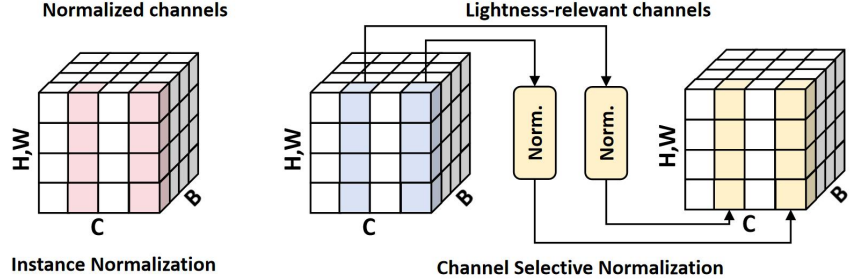

**Fig. 3.** CSNorm selectively normalizes brightness-related channels to enhance generalization, while leaving other channels unchanged for precise reconstruction.

### Feature Consistency Module (FCM)

Feature Consistency module contains a convolution layer with $3 \times 3$ filters, a ReLU activation and a Batch Normalization, which can be denoted as follows:

$$FCM(\cdot) = BN \circ LReLU \circ Conv_{3\times3}(\cdot) \tag{4}$$

Further elaborating on the system's architecture, the condition network is tasked with generating a condition vector $V$, critical for the adaptive mapping process. The network's structure is adeptly designed to accommodate a down-sampled SDR image as its input, subsequently producing the condition vector through the following formulation:

$$V = GAP \circ Conv_{1\times1} \circ Dropout \circ Conv_{1\times1} \circ Concat(CHM^{N_c}(M_S), FCM^{N_c}(M_S)) \tag{5}$$

where $GAP$ denotes global average pooling, and $Dropout$ is applied as a regularization technique to mitigate overfitting risks by introducing multiplicative Bernoulli noise to the features, akin to the effect of data augmentation.

A distinctive characteristic of the condition network, facilitated by the exclusive use of 1×1 convolutional filters, is its focus on global rather than local feature extraction. This is complemented by pooling layers that enable the extraction of global priors grounded on image statistics, emphasizing the network's capacity to discern and leverage overarching color patterns and trends within the image. Intriguingly, the network's design allows for robust performance even under conditions of pixel shuffling, underscoring the notion that the pivotal priors for global color mapping are largely independent of spatial arrangements within the image. This insight further elucidates the network's adeptness at abstracting essential color mapping cues from a global perspective, reinforcing the efficacy of the proposed FastHDRNet in achieving high-fidelity SDR-to-HDR translation.

### Global Feature Modulation (GFM)

To further enrich the feature modulation within the base network, we adopt the GFM strategy [8], a technique heralded for its efficacy in photo retouching applications.

GFM modulates the intermediate feature maps via scaling and shifting operations, mathematically described as:

$$GFM(x_i) = \alpha_1 * x_i + \alpha_2 \tag{6}$$

where $x_i$ denotes the intermediate feature map, and $\alpha_1, \alpha_2$ are the modulation parameters for scaling and shifting, respectively.

The aggregate operation of the AUCT network is thus formalized as:

$$M_{AUCT} = GFM \circ Conv_{1 \times 1} \circ (ReLU \circ GFM \circ Conv_{1 \times 1})^{N_l - 1}(M_S) \tag{7}$$

where $M_{AUCT}$ represents the resultant output image. To ensure the fidelity of the AUCT output to the target HDR image, an optimization objective is employed, minimizing the L2 norm between the AUCT output and the ground truth HDR image, thereby guiding the network towards a precise emulation of HDR imagery.

### 3.2 Local Enhancement (LE)

Following the AUCT phase in our SDR-to-HDR conversion pipeline, we implement a LE stage to refine the visual quality further. Despite the significant improvements rendered by AUCT, the incorporation of LE is critical for augmenting the fidelity of the resultant HDR images. It has been empirically observed that applying local operations directly for end-to-end mapping, prior to AUCT, frequently results in pronounced artifacts in the output, underscoring the necessity of a sequential approach. Detailed insights into these observations are provided in the supplementary material.

For the LE stage, we employ a U-Net architecture [41] due to its proven efficacy in feature refinement. While the exploration of more sophisticated architectures remains within the scope of future work, the current focus is on demonstrating the utility of U-Net within our proposed framework. The LE process is divided into the main branch and condition branch. The main branch is a U-shape structure and the condition branch generates the condition vector to modulate the intermediate features of the main branch. We take the output of $M_{AUCT} \in R^{3 \times H \times W}$ as input for the local enhancement network. In the main branch , the input is transformed to a high-dimensional feature $F \in R^{C \times H \times W}$, where C is the number of channels. Then this feature goes through a U-Net which is a three-level encoder-decoder structure. In real-world scenarios, SDR content is typically in the range of 1K to 4K resolution. The use of a U shape structure can significantly reduce the computational cost and fasten the inference time. As for the condition branch, we use three convolutions to produce three different -size features, which can be used to modulate the intermediate feature of the main branch. In the part of LE, we aim to implement the region-dependent operation and further address the spatially variant mapping for SDR-to-HDR. Thus, we employ the SFT layer [29] and leverage its ability of spatial feature modulation to build the network. The SFT layer is denoted as:

$$SFT(x_i) = m \circ x_i + n \tag{8}$$

where ∘ denotes the element-wise multiplication. $x_i \in R^{C \times H \times W}$ is the intermediate features to be modulated. $R \in R^{C \times H \times W}$ and. $n \in R^{C \times H \times W}$ are two condition maps predicted by the condition branch.

This methodology underscores our commitment to improving the quality of SDR-to-HDR conversion, leveraging both global and local processing strategies to achieve an output that closely mirrors the realism and detail of high dynamic range content.

And the whole pipeline is as followed:

$$M_H = LE(AUCT(M_s)) \tag{9}$$

As for the loss function we use here is:

$$L_{HR} = \alpha L_1 \tag{10}$$

where $\alpha$ is the loss weight and we set it to 0.00001.

## 4 Experiments

### 4.1 Experimental Setup

**HDRTV1K Dataset Construction**
Addressing the challenge of the limited availability of paired SDR/HDR datasets for training and validation, we developed the HDRTV1K dataset, adhering to the HDR10 standard [5]. This dataset encompasses 22 HDR video sequences and their corresponding SDR versions, encoded in PQ-OETF within the Rec.2020 color space. For experimental validation, 18 pairs were designated for training, with the remaining 4 pairs set aside for testing. By extracting one frame every two seconds from each video, we compiled a training set of 1235 images and a test set of 117 images, ensuring content diversity and minimizing scene repetition.

**Training Protocol**
For the proposed AUCT, the base network consists of 3 convolution layers with 1×1 kernel size and 128 channels, and the condition network contains 4 CHMS and 4 FCMS. Before training, we crop images by 480 × 480. During training, patches of size 480 × 480 are input into the base network, while full images downsampled by a factor of 4 are input into the condition network. We set the mini-batch size to 16 and use the L1 loss function and Adam optimizer for training, with a total of $1 \times 10^6$ iterations. The initial learning rate is set to $4 \times 10^{-6}$, and is decayed by a factor of 4 at the $5 \times 10^5$ iterations . And we train the AUCT and LE networks together to achieve better effects.

**Evaluation Metrics**
To rigorously evaluate our SDR-to-HDR conversion method, we utilize five key metrics for a thorough comparison: Peak Signal-to-Noise Ratio (PSNR), Structural

Similarity Index Measure (SSIM), Spectral Residual-Based Similarity (SR-SIM) [38], High Dynamic Range Visual Difference Predictor version 3 (HDR-VDP3) [40], and ΔE_ITP [39]. PSNR assesses image fidelity, while SSIM and SR-SIM, with SR-SIM's effectiveness in HDR proven by measure image similarity. ΔE_ITP, tailored for HDR, quantifies color differences accurately. HDR-VDP3, updated for Rec.2020 color space, provides a detailed image quality analysis, with settings that include a "side-by-side" comparison, "rgb-bt.2020" encoding, 50 pixels per degree visual acuity, and "led-lcd-wcg" display, aligning perfectly with our evaluation framework.

**Visualization Approach**

Our approach to displaying HDR images in 16-bit PNG format, rendered on SDR screens via gamma EOTF, might lead to perceived brightness differences. However, our methodology preserves clear visual distinctions. Contrary to prior methods [14], [15] that employed media players for HDR visualization—potentially introducing biases due to player-specific enhancements—our technique aligns more closely with natural visual perception. By circumventing error map visualizations that fail to accurately convey perceptual differences and maintaining detail in highlighted areas, our approach ensures visual contrasts are in close alignment with human observational capabilities, as demonstrated in **Fig. 4**.

**4.2     Comparison with Existing Methods**

In assessing our method's effectiveness, we benchmarked it against various techniques, including joint Super-Resolution (SR) and SDR-to-HDR conversion, as well as image-to-image translation, photo retouching, and LDR-to-HDR conversion, with results consolidated in **Table 1**.

**Quantitative Evaluation**

Our framework excelled in all metrics, showcasing the LE network's integration into our AUCT as a superior approach, as documented in **Table 1**. The PSNR metric notably achieved 37.67dB, reducing parameter complexity.

**Visual Comparison Analysis**

After conducting a detailed analysis of visual results across various methodologies, as illustrated in **Fig. 4**, it has been observed that LDR-to-HDR conversion and image-to-image translation methods generally produces images with reduced contrast. Additionally, apart from HuoPhyEO [36], these methods often yield images with unnatural colors and evident artifacts. Although photo retouching techniques show slight improvements, they still suffer from color bias issues.

Contrarily, our approach stands out by producing images with authentic color fidelity and improved contrast, accurately reflecting the ground truth without introducing unnecessary artifacts. By methodically applying AUCT followed by LE, our framework enhances the visual quality of outputs. This stepwise improvement demonstrates the effectiveness of our comprehensive approach in navigating the complexities of SDR-to-HDR conversion.

**Efficiency Analysis**

In **Table 2**, we provide the comparison of computational complexity: Multiply－Accumulate Operations (MACs) and the average inference time between our DFT and the previous state-of-the-art method HDRTVNet [5]. Each input image in the HDRTV1K has a spatial size of $3840 \times 2160$. It can be seen that our **FastHDRNet** significantly reduces the computational cost and runs much faster compared to HDRTVNet while achieving better performance.

**Table 1.** Quantitative comparisons with existing methods.

| Method | | Params↓ | PSNR↑ | SSIM↑ | SR-SIM↑ | $\Delta E_{\text{ITP}}$ ↓ | HDR-VDP3↑ |
|---|---|---|---|---|---|---|---|
| Photo Retouching | HDRNet[30] | 482K | 35.73 | 0.9664 | 0.9957 | 11.52 | 8.462 |
| | CSRNet[31] | 36K | 35.04 | 0.9625 | 0.9955 | 14.28 | 8.400 |
| | Ada-3DLUT[32] | 594K | 36.22 | 0.9658 | 0.9967 | 10.89 | 8.423 |
| Image-to-image translation | ResNet[33] | 1.37M | 37.32 | 0.9720 | 0.9950 | 9.02 | 8.391 |
| | Pixel2Pixel[34] | 11.38M | 25.80 | 0.8777 | 0.9871 | 44.25 | 7.136 |
| | CycleGAN[35] | 11.38M | 21.33 | 0.8496 | 0.9595 | 77.74 | 6.941 |
| SDR-to-HDR | JSI-GAN[15] | 1.06M | 37.01 | 0.9694 | 0.9928 | 9.36 | 8.169 |
| | Deep SR-ITM[14] | 2.87M | 37.10 | 0.9686 | 0.9950 | 9.24 | 8.233 |
| HDRTVNet[5] | Base Network | 5K | 36.14 | 0.9643 | 0.9961 | 10.43 | 8.305 |
| | AGCM | 35K | 36.88 | 0.9655 | 0.9964 | 9.78 | 8.464 |
| | AGCM-LE | 1.41M | 37.61 | 0.9726 | 0.9967 | 8.89 | 8.613 |
| | AGCM-LE-HG | 37.20M | 37.21 | 0.9699 | 0.9968 | 9.11 | 8.569 |
| LDR-to-HDR | HuoPhyEO[36] | - | 25.90 | 0.9296 | 0.9881 | 38.06 | 7.893 |
| | KovaleskiEO[37] | - | 27.89 | 0.9273 | 0.9809 | 28.00 | 7.431 |
| **FastHDRNet (ours)** | AUCT-LE | 514K | 37.67 | 0.9710 | 0.9968 | 8.60 | 8.620 |

1 Red, blue and green texts indicate the best, second best and third best results.

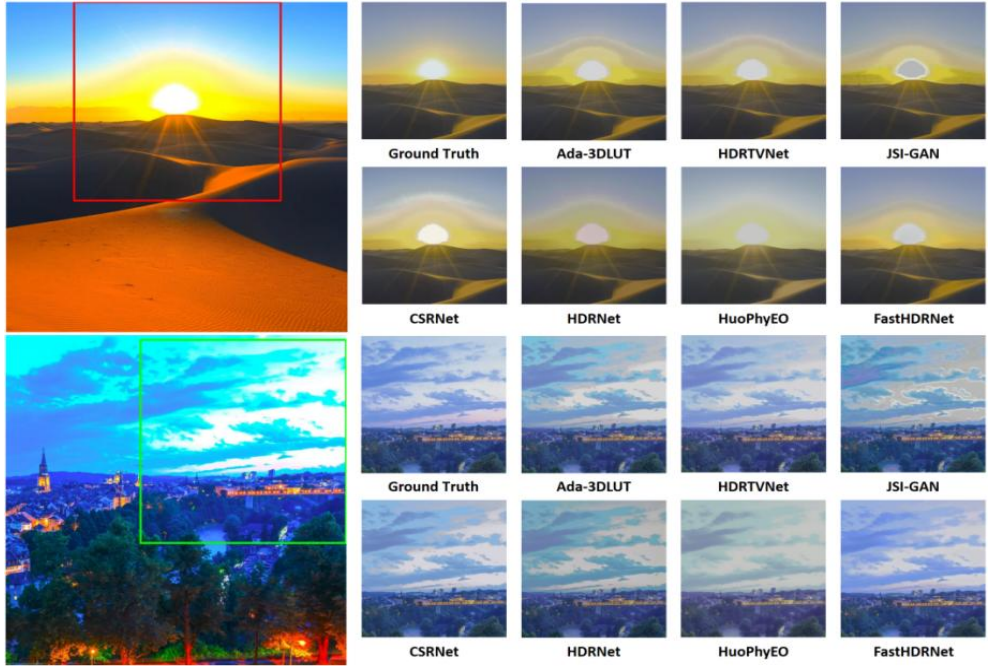

**Fig. 4.** Visual comparison on real dataset.

**Table 2.** Comparisons of MACs, inference time and PSNR between HDRTVNET and our FastHDRNet

| Method | MACs(G)↓ | Time(s)↓ | PSNR↑ |
|---|---|---|---|
| HDRTVNET[5] | 3112.99 | 0.24 | 37.61 |
| **FastHDRNet(ours)** | **502.78** | **0.1** | **37.67** |

### 4.3    User Study

To subjectively assess the visual quality of HDRTVNet, a user study was conducted with 40 participants, comparing it against the top-performing methods across various categories. For this purpose, 30 images were randomly selected from the test set and showcased on a HDR television (LG 32GQ950, with peak brightness of 1000 nits) within a controlled darkroom setting. Prior to the commencement of the study, participants were briefed on evaluating the images based on three criteria: the presence of obvious artifacts and unnatural colors, the naturalness and comfort of color, brightness, and contrast, and the delineation of contrast between light and dark areas, including the rendering of highlight details. Participants were then asked to rank the results of each method according to these criteria. The television was calibrated to the Rec.2020 color gamut and HDR10 standard for displaying the results.

The methods under comparison included Ada-3DLUT[32], JSI-GAN[15], CSRNet [31], HDRTVNet[5], and FastHDRNet, juxtaposed with the ground truth. FastHDRNet was respectively considered to provide the best visual quality in 42.8% (513 counts) of instances, with FastHDRNet also securing 30.9% (371 counts) for the second-best visual quality ranking.

### 4.4 Ablation Study

As you can see from **Table 3**, the integration of LE part into AUCT shows great effects, which elevate the PSNR from 36.95 to 37.62. The improvement demonstrates the significant role that the LE component plays in enhancing the performance of AUCT algorithm. This increase is not only statistically significant but also visually apparent in the improved quality of the color-transformed images.

**Table 3.** Ablation Study

| Method | PSNR↑ |
|---------|-------|
| AUCT | 36.95 |
| AUCT-LE | 37.62 |

## 5 Conclusion

This research delves into the transformative journey from SDR to HDR television, highlighting a significant advancement in visual quality enhancement within the television production sphere. We introduced FastHDRNet, a novel cascaded framework employing two specialized deep neural networks tailored for the SDR-to-HDR conversion process. Our method, incorporating pixel-independent operations that account for both average and peak brightness and region-dependent refinements via region-independent operations, effectively minimizes the introduction of artifacts compared to conventional methods.

Our work offers a substantial contribution by providing a lightweight and efficient SDR-to-HDR conversion solution with leading performance. Through a new normalization method, our network's robustness is improved, positioning our approach as pivotal in HDR technology advancement. Our method stands out for its swift reference time among evaluated algorithms, demonstrating our contributions' potential to redefine standards in HDR content production and display quality. The advancement of such research is expected to enhance consumer visual experience and facilitate the wider adoption of HDR technology in the media and entertainment industries.